\documentclass[11pt,a4paper]{article}
\usepackage{jheppub}

\title{Remarks on the Confinement in the $G(2)$ Gauge Theory Using the Thick Center Vortex Model } 
\author[a]{H.~Lookzadeh} 
\affiliation[a]{Faculty of Physics, Yazd University,\\ P.O. Box 89195-741, Yazd, Iran}
\emailAdd{h.lookzadeh@yazd.ac.ir}

\abstract{The confinement problem is studied using the thick center vortex model. It is shown that the $SU(3)$ Cartan sub algebra of the decomposed $G(2)$ gauge theory can play an important role in the confinement. The Casimir eigenvalues and ratios of the $G(2)$ representations are obtained using it’s decomposition to the $SU(3)$ subgroups. This leads to the conjecture that the $SU(3)$ subgroups also can explain the $G(2)$ properties of the confinement. The thick center vortex model for the $SU(3)$ subgroups of the $G(2)$ gauge theory is applied without the domain modification. Instead, the presence of two $SU(3)$ vortices with opposite fluxes due to the possibility of decomposition of the $G(2)$ Cartan sub algebra to the $SU(3)$ groups can explain the properties of the confinement of the $G(2)$ group both at intermediate and asymptotic distances which is studied here.}
\begin{document}
\maketitle
\flushbottom

\section{Introduction}
Quantum Chromo dynamics (QCD) at low energies is dominated by non-perturbative phenomena of the quark confinement and spontaneous chiral symmetry breaking (SCSB). Quarks and gluons as the building blocks of a gauge theory for the strong interaction are not present in the QCD spectrum. This leads to the confinement phenomena \cite{Greensite2011}. Now a days from lattice gauge theory simulations it is known that the confinement phenomenon is present in a non abelian gauge theory \cite{ wilson1974, creutz1980,creutz1983, Rothe1983, Montvay}. 
 So we must search for a model or mechanism to satisfy the confinement problem in a non abelian gauge theory. For example a super symmetric theory is not essential for the confinement \cite{seiberg1994}.

A good model or theory of the confinement should obey all of the properties of the confinement which are present in the lattice gauge theory calculation. One way to study the confinement problems in a theory is to obtain the potential part of the interaction energy between the static sources. The kinetic part of energy can be eliminated if the quark is studied when they are massive \cite{M. A. Shifman1999,Bali2001}. The properties of the confinement in a gauge theory can be described by studying the static potential. According to the Regge trajectories from experimental data \cite{Regge1959, Regge1960,Chew1961,Chew1962} and also the results from the lattice gauge theory \cite{creutz1983,Rothe1983, Montvay,Greensite2011} , phenomenological models for the quark confinement are introduced \cite{Hooft1978, Hooft1978, Greensite2011,Bali2001} . In these models, the QCD vacuum is filled with some topological configurations that confine the colored objects. The most popular candidates among these topological fields are the monopoles and the vortices. Other candidates include the instantons, merons, calorons, dyons. However Greensite shows that only the vortex model could obey a difference in-area-law \cite{greensite2015} and other models such as the monopoles gas, caloron ensemble, or the dual abelian Higgs actions cannot obey the law. Each of these defects have some advantages relative to the others and the confinement problem can be studied by these models. For example the calaron can explain the relation of the confinement to the temperature or the thick center vortex model can explain the Casimir scaling. In this article the thick center vortex model is used to study the confinement problem\cite{Faber1998}.

The center vortex model was initially introduced by 't Hooft\cite{Hooft1978}. It is able to explain the confinement of quark pairs at the asymptotic region, but it is not able to explain the confinement at intermediate distances especially for the higher representations. Following 't Hooft, the idea that thick center vortices are relevant for the confinement mechanism has been introduced by Mack and Petkova \cite{Mack1979, Mack1980, Mack19801, Mack1982}. The model then is modified to the thick center vortex model by Greensite, Faber, etc\cite{Faber1998}. Within this model one can obtain the Casimir scaling and N-ality behavior of the gauge theory. This is done before for the $SU(2)$, $SU(3)$ and $SU(4)$ groups \cite{Deldar2005,Rafibakhsh2007,Rafibakhsh2010}. Using another modification of the model to the domain vacuum structures it is possible to describe the properties of a gauge theory without non trivial center such as the $G(2)$ gauge theory\cite{Greensite2007, Lookzadeh2012,Deldar2014, Rafibakhsh2014, Deldar2015} and also a better Casimir scaling.

In this article it is tried to show that the $G(2)$ group has the $SU(3)$ decomposed subgroups which can explain the confinment properties at intermediate and asymptotic distances without modification to the domain structure. In the next section the $G(2)$ group properties are introduced especially the topological properties for the presence of the topological solitonic structures. In the section $III$ the thick center vortex model is introduced and then the Casimir ratios of the $G(2)$ are obtained analytically. Obtaining the $G(2)$ Casimir ratios exactly with the use of its decomposition to the $SU(3)$ subgroup leads to the conjecture that any properties of the confinement may be dominated by its $SU(3)$ subgroups. To study this issue the behavior of the confinement in the $G(2)$ gauge theory is obtained in the section $IV$ by applying the thick center vortex model to the $SU(3)$ subgroups of the $G(2)$ group. In the section $V$ the properties of such "internal vortices" are explained.
\section{The $G(2)$ Group Properties}
 The exceptional Lie group $ G(2)$ is the auto morphism group of the octonion algebra \cite{Georgi,MACFARLANE2001, Pepe2003, Pepe2007, Pepe2011, Bjorn2011}. The group $G(2)$ is a simply connected, compact group. The $G(2)$ is its own covering group and its center is trivial. As it is clear the rank of  the $G(2)$ is $2$ and it has $14$ generators. In the Cartan sub algebra which is the representation which the most simultaneous diagonal generators for the generators of a group is present, the $G(2)$ has two diagonal generators and the remaining $12$ generators are not diagonal. Since there are $14$ generators the adjoint representation of $G(2)$ can be introduced by $14 \times 14$ matrices. So it is a $14$ dimensional real group. Also its fundamental representation is $7$ dimension. Since it is the subgroup of the $SO(7)$ with rank $3$ and $21$ generators, its elements can be obtained by the $SO(7)$ elements obeyed the $7$ constrained reduced to $14$ generators of the fundamental representation. If the $Us$, the $7 \times 7$ real orthogonal matrices with determinant $1$ is considered, then
 
 \begin{eqnarray}
UU^{\dagger}=1,
\end{eqnarray} 
is the constrained matrices that are elements of the $SO(7)$. Within the constrained called cubic constrained which is
\begin{eqnarray}
T_{abc}=T_{def}U_{da}U_{eb}U_{fc},
\end{eqnarray}
and $T$ is totally antisymmetric tensor, and its nonzero elements are 
\begin{equation}
T_{127}=T_{154}=T_{163}=T_{235}=T_{264}=T_{374}=T_{567}=1,
\end{equation}
the 14 number of generators of the $G(2)$ are obtained and the $Us$ become the $G(2)$ elements. 
As the Cartan sub algebra for the calculation here should be used, the two diagonal generators of its fundamental representation are
\begin{equation}
H^3=\frac{1}{\sqrt{8}}(p_{11}-p_{22}-p_{55}+p_{66}),                             \\
H^8=\frac{1}{\sqrt{24}}(p_{11}+p_{22}-2p_{33}-p_{55}-p_{66}+2p_{77}),
\end{equation}
Where $(p_{ij})_{\alpha \beta}=\delta_{i\alpha} \delta_{j\beta}$ and $\alpha, \beta$ indicate the row and the column of the matrices, respectively. These two diagonal generators can be builded with the $SU(3)$ diagonal Cartan sub algebra generators of $SU(3)$ such as 
\begin{eqnarray}
H^a=\frac{1}{\sqrt{2}}diagonal(\lambda^{a},0,-(\lambda^{a})^*),
\end{eqnarray}  
$\lambda^a (a=3,8)$ are the two diagonal Cartan generators of the $SU(3)$. It is not possible to construct $H^8$ from the $SU(2)$ diagonal Cartan sub algebra. 

As we are interested in the defect structures in a gauge theory, the topological properties of a group is important for us \cite{Nakahara,Reinhardt2002, Maas2012}. The $G(2)$ manifold is a seven-dimensional Riemannian manifold with holonomy group contained in the $G(2)$. Its first fundamental group is trivial 
\begin{eqnarray}
\pi_{1}(G_{2})=I.
\end{eqnarray}  
This shows that there is not any vortex defect present within a $G(2)$ gauge theory. As the $SU(3)$ is a subgroup of the $G(2)$ any element of the $G(2)$ such as $U$ can be written as 
\begin{eqnarray}
U=S.V, V\in SU(3) \text{  and  } S\in \frac{G(2)}{SU(3)}\sim S^{6}.
\end{eqnarray}  
However 
\begin{eqnarray}
\pi_{1}(SU(3))=I \text{  and  } \pi_{1}(S^6)=I.
\end{eqnarray}  
This shows that one cannot find any vortex structures within $G(2)$ subgroup without any symmetry breaking or gauge fixing or singular transformation. The center element of $G(2)$ is trivial, so a center transformation leads to 
\begin{eqnarray}
\pi_{1}(\frac{G(2)}{I})=I.
\end{eqnarray}  
Again no vortex structure but for its $SU(3)$ subgroup under center transformation it leads to 
\begin{eqnarray}
\pi_{1}(\frac{SU(3)}{Z_{3}})=Z_{3}. 
\end{eqnarray}    
This shows a center vortex is present in $SU(3)$ part of $G(2)$ elements $U$ after center transformation. So despite the gauge group does not have a center vortex with regard to its center it possess a vortex with regard to its subgroup center transformation 
\begin{eqnarray}
\pi_{1}(\frac{SU(3)\times S^6}{Z_{3}})=Z_{3}.
\end{eqnarray}    
Also we have 
\begin{eqnarray}
\pi_{2}(G(2))=I.
\end{eqnarray}  
So there is no monopole structure within this group. But after an spontaneously symmetry breaking to $U(1) \times U(1)$ special type of monopole emerges \cite{shnir2015}
\begin{eqnarray}
\pi_{2}(\frac{G(2)}{U(1)\times U(1)})=Z.
\end{eqnarray}
The symmetry breaking of the $G(2)$ group to the residual subgroup also leads to \cite{Di Giacomo}:
\begin{eqnarray}
\pi_{2}(\frac{G(2)}{SU(2)\times U(1)})=Z.
\end{eqnarray}
Which again leads to the monopoles structures. Also the fundemental group of rank $3$ of the $G(2)$ is not trivial
\begin{eqnarray}
\pi_{3}(G(2))\neq I.
\end{eqnarray}
    
It leads to the presence of the instanton structure within this group. Despite there are five exceptional group which are $G(2),F(4),E(6),E(7),E(8)$ only $G(2),F(4),E(8)$ have trivial center elements. The $E(6)$ center is $Z(3)$ and the $E(7)$ center is $Z(2)$. The $G(2)$ is the simplest exceptional group among these groups. Two properties of trivial center element and being its own corvering group leads the $G(2)$ gauge theory interesting to study the properties of the confinement.

\section{The Thick Center Vortex and Casimir Scaling}
The idea of the vortex model for the confinement is due to 't Hooft and Mandelstam \cite{Mandelstam1976, Hooft1978, Hooft1979, tHooft1981, tHooft1982}. They used the Nielson Oleson vortex solution to obtain the confinement properties \cite{Nielson1973}. In the dual superconductivity the string between the sources of abelian electric charges is due to the abelian magnetic charge condensation and the string obey the vortex type equations (for example the string between quark-antiquark in a meson). In the center vortex picture presence of the vortices in the vacuum is due to the center elements and their fluctuation in the number of center vortices linked to the Wilson loop leads to an area law Wilson loop and a linear potential or string like behavior. In the dual superconductor the vortices have electric flux, but in the center vortex picture the vorices have magnetic flux. Kronfeld and etc suggested test 't Hooft theory in the lattice gauge theory\cite{Kronfeld1987}. The idea that thick center vortices are relevant for the confinement mechanism has been introduced by Mack and Petkova \cite{Mack1979, Mack1980, Mack19801, Mack1982}. The thick center vortex model has been introduced by Del Debbio, Faber, J. Greensite and Olejnik to obtain the intermediate behavior of the quark potential using the lattice gauge theory (LGT) results \cite{Del Debio1998}.

The center vortex is a topological field configuration which is line like in D=3 dimensional and surface like in D=4 dimension and have some finite thickness. A discontinuity in the background gauge transformation related to the gauge group center leads to a center vortex. The center vortex creation linked to a Wilson loop, in the fundamental representation of SU(N) changed the Wilson loop holonomy by an element of the gauge group center

\begin{equation}\label{eqwilson} 
W(C)\longrightarrow e^{\frac{2\pi n i}{N}} W_{0}(C),
\end{equation} 
The confinement is obtained from random fluctuations in the linking number. A vortex piercing a Wilson loop contribute with a center element $Z$ somewhere between the group elements of the gauge group 

\begin{equation}\label{eqwilson1} 
W(C)=Tr[UUU...U ]\longrightarrow Tr[UU....(Z)U].
\end{equation} 
The center elements commute with all members of the group, so the location of $Z$ in eq. \ref{eqwilson1} can be changed by changing the place of discontinuity which leads to a vortex formation. An equation for the string tension $\sigma$ can be obtained assuming that the vortices are thin and pierce Wilson loops in a single plaquette with the independent probability $f$. For example for the $SU(2)$ group the following is obtained 

\begin{eqnarray}\label{eqwilson2}
\langle W(C)\rangle =\prod \lbrace (1-f)+f(-1) \rbrace \langle W_{0}(C)\rangle =exp[-\sigma(C) A] \langle W_{0}(C)\rangle.
\end{eqnarray} 

$\langle W_{0}(C)\rangle$ is the expectation value of the loop when no vortex pierces this loop. $A$ is the area of the Wilson loop and is equal to $R\times T$. $R$ is for the space side of the Wilson loop and $T$ is for the time side. Then for the string tension it leads to 

\begin{equation}\label{eqsigma} 
\sigma =\frac{-1}{A} ln(1-2f).
\end{equation} 
The vortex model works very well for the fundamental representation and the adjoint at the large distances. Lattice simulations show an intermediate string tension for the higher representation \cite{Bali2000,Deldar2000,Bernard1982,Bernard1983,Ambjørn1984,Michael1985} which cannot be introduced by the thin center vortex model. The thick center vortex has been introduced to obey the intermediate string tension by Faber, Greensite and Olejnik \cite{Faber1998}. The lattice data shows that vortices have comparable thickness \cite{Mack1979, Mack1980, Mack19801, Mack1982, Faber1998, Del Debio1998}. So the thickness of the vortex must be calculated in the 't Hooft model. The first assumption of the thick center vortex is to consider a gauge group element $G$ instead of $Z$. So if a thickness is present, the vortex piercing to a Wilson loop can be described by an element of the gauge group $G$ instead of $Z$ 

\begin{equation}\label{eqwilson3} 
W(C)=Tr[UUU...U ]\longrightarrow Tr[UU....(G)U].
\end{equation}
$G$ is a group factor and can gets the values between trivial element and non-trivial center element of the gauge group. $U$s are the group elements and get their value as 

\begin{equation}\label{theta} 
U=e^{i n^{a}T_{a}}
\end{equation}
where the $n^a$s introduce the space of the manifold of the group and $T_a$s are the generators of the group in any representation and $a$ is the number of the generators. But $G$ is 
\begin{eqnarray}\label{groupfactor} 
G(x,S)=exp[i\alpha_{c}(x)\vec{n}.\vec{T}]=
S exp[i\alpha_{c}(x)\vec{n}.\vec{H}] S^{\dagger}.
\end{eqnarray}
The $H_i$ are the generators that span the Cartan sub algebra in the $r$-representation. $S$ is the gauge group element in the $r$-
representation. $\alpha_{c}(x)$ relates to the vortex center $x$ relative to the Wilson loop $c$. 

The second assumption for the thick center vortex \cite{Faber1998} is that the probabilities $ f_n $ that plaquette in the minimal area are pierced by the vortex of type $n$ are uncorrelated. The random color group orientation associated with $S$ are also uncorrelated and should be averaged. This is an over simplification of vortex thickness effects but at least provide an acceptable picture of vortex thickness. To do this the color group manifold should be averaged over. This leads to
  \begin{eqnarray}\label{groupfactor2} 
\bar{G}(\alpha)=\int dS S exp[i\alpha_{c}(x)\vec{n}.\vec{H}] S^{\dagger}=g_{r}(\alpha) I_{d_r},\qquad
g_{r}(\alpha)=\frac{1}{d_r}Tr exp[i\vec{\alpha}.\vec{H}].
\end{eqnarray}
The $\vec{H}$ generators are used from Cartan subalgebra because in this algebra the most simultaneous commutable generators with other generators are available. $I_{d_r}$ is a unit matrix with the dimension of $r$. Only the effects of the abelian parts of the color degrees of freedom is considered on the vortex thickness. The most abelian and diagonal generators are present in the Cartan sub algebra \cite{Georgi}. By this assumption the calculations are restricted to the Cartan sub algebra for the application of the thick center vortex. The $\alpha_{c}(x)$ is a profile ansatz to consider the portion of the center vortex on a Wilson loop and $\alpha_{c}(x)$ must obey all the boundary criteria that can be used as an ansatz. The most available ansatzs previously are the one introduced in \cite{Faber1998} and the other one introduced in \cite{Greensite2007}. 

The $G(x,\alpha)$ is used instead of $Z$ to show the effect of thickness of center vortices. So it should obey all the portion of vortex intersection with the Wilson loop. So it must has an abelian character. Also the model must obey the Casimir scaling, so the projection on the $H_i$ of the group is used. The question here is what would happens if the Cartan sub algebra within this model is used? To answer this question the Casimir scaling can be obtained analytically. It is known that 
\begin{equation}\label{generator} 
\frac{1}{d_r}Tr(H_{i}H_{j})=\frac{C_{r}^{(2)}}{N^2-1}\delta_{ij},
\end{equation}  
Which $H_{i}$s are the Cartan generators. Here the diagonal Cartan generators are used due to the commutation properties of the $G$ with $U$s. The $C_{r}^{(2)}$ is quadratic Casimir eigenvalue of representation $r$. Also if the lowest series expansion are kept, 
\begin{equation}\label{generatorexpansion} 
\frac{1}{d_r}e^{i\vec{\alpha}.\vec{H}}\cong \frac{1}{d_r}Tr(1+i\vec{\alpha}.\vec{H}+\frac{1}{2}\alpha_{i}\alpha_{j}H_{i}H_{j}+...),
\end{equation}
\begin{equation}\label{generatorexpansion1}
\frac{1}{d_r}Re(e^{i\vec{\alpha}.\vec{H}}) \cong \frac{1}{d_r} Tr(1+\frac{1}{2}\alpha_{i}\alpha_{j}H_{i}H_{j}+...).\nonumber
\end{equation}
then the string tension of the potential is obtained as the following

\begin{equation}\label{stringexpansion} 
\sigma_{c}=-\tfrac{1}{A}\sum_{x}(1-(1-\sum_{n=0}^{N-1} f_{n}(-Tr(\frac{1}{2}\alpha_{i}\alpha_{j}H_{i}H_{j}))
=\frac{1}{A}\sum_{x} \sum_{n=1}^{N-1}\frac{f_n}{2(N^2-1)}\vec{\alpha}_{c}^{n}(x).\vec{\alpha}_{c}^{n}(x) c_{r}^{(2)}
\end{equation} 
 
\begin{figure}[h!]
\begin{center}
\resizebox{0.6 \textwidth}{!}{\includegraphics{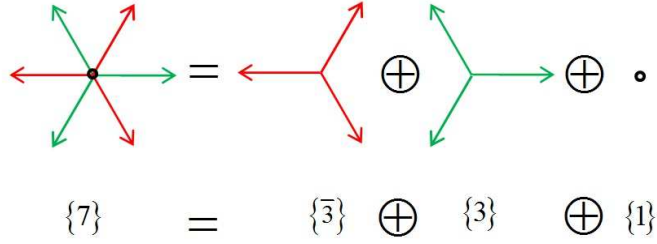}}

\caption{
The weight diagram of the $G(2)$ group and its decomposition to the $SU(3)$ group in the weight space.}
\label{weight7}
\end{center}
\end{figure}

If $\alpha_{c}(x)=constant$, then the Casimir scaling is obtained but $\alpha_{c}(x)$ depends on the loop size and is not a constant. This shows that if the Cartan subalgebra of a gauge symmetry is used, obtaining the Casimir scaling analytically is possible within the thick center vortex model. Now consider this for the $G(2)$ gauge group using its decomposition. Consider decomposition in to the $SU(3)$ groups. The $G(2)$ weight diagram for fundamental representation is shown in figure \ref{weight7}. Also its decomposition to the $SU(3)$ subgroup is shown. Because it is possible to explain the weight vector of the $G(2)$ using the $SU(3)$ weight vector, it leads to
\begin{equation}\label{sevendecomposition} 
\lbrace 7 \rbrace=\lbrace 3\rbrace\oplus \lbrace\bar{3}\rbrace \oplus\lbrace 1\rbrace ,
\end{equation} 
then  
\begin{equation}\label{cartandecomposition} 
H_{\lbrace 7 \rbrace}=H_{\lbrace 3\rbrace}\oplus H_{\lbrace\bar{3}\rbrace} \oplus H_{\lbrace 1\rbrace}.
\end{equation}   
Multiplying the two Cartan generators 
\begin{multline}\label{cartanproduct} 
H_{\lbrace 7 \rbrace i}\otimes H_{\lbrace 7 \rbrace j}=
\lbrace H_{\lbrace 3\rbrace i}\oplus H_{\lbrace\bar{3}\rbrace i} \oplus H_{\lbrace 1\rbrace i} \rbrace  \otimes \lbrace H_{\lbrace 3\rbrace j}\oplus H_{\lbrace\bar{3}\rbrace j} \oplus H_{\lbrace 1\rbrace j} \rbrace .
\end{multline}
The trace of the above expression is needed. So only the acceptable matrix products are kept 
 \begin{equation}\label{cartanproduct2} 
Tr \lbrace H_{\lbrace 7 \rbrace i}\otimes H_{\lbrace 7 \rbrace j}\rbrace =Tr\lbrace H_{3i}H_{3j} \oplus H_{\bar{3}i}H_{\bar{3}j}\oplus
H_{1i}H_{1j}\rbrace ,
\end{equation}
Using the eq. \ref{generator} it leads to
\begin{multline}\label{cartanproduct3}
Tr \lbrace H_{\lbrace 7 \rbrace i} \otimes H_{\lbrace 7 \rbrace j}\rbrace =2C^{2}_{\lbrace 3\rbrace}\frac{d_r}{N^2-1}+0 
=2\frac{(N^2-1) d_{r}}{2N^2}\frac{d_{r}}{N^2-1}+0=1,
\end{multline}

\begin{figure*}
\begin{center}
\resizebox{0.6\textwidth}{!}{
\includegraphics{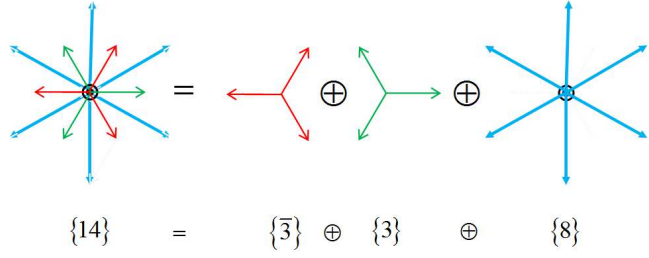}}

\caption{\label{root7}
The root diagram of the $G(2)$ group and its decomposition to the $SU(3)$ group in the root space.}
\end{center}
\end{figure*}
where $d_r$ is the dimension of the representaion of the $SU(3)$ subgroup and $N$ is the dimension of the subgroup.  
The eq. \ref{cartanproduct3} is exactly the Casimir eigen value for the $G(2)$ in fundamental representation. So the Casimir eigen value of the $G(2)$ using its decomposition to the $SU(3)$ subgroup is obtained. Here the same is done for the adjoint representation of the $G(2)$ group. The $G(2)$ root diagram and its decomposition to its $SU(3)$ subgroup is shown in figure \ref{root7}.   
Again the $SU(3)$ decomposition is possible: 
  \begin{equation}\label{14decomposition} 
\lbrace 14 \rbrace=\lbrace 8\rbrace\oplus \lbrace 3\rbrace \oplus\lbrace \bar{3}\rbrace,
\end{equation} 
and then
\begin{equation}\label{14cartandecomposition} 
H_{\lbrace 14 \rbrace}=H_{\lbrace 8\rbrace}\oplus H_{\lbrace 3\rbrace} \oplus H_{\lbrace \bar{3}\rbrace},
\end{equation} 
multiplying the two Cartan Generators
\begin{multline}\label{14cartanproduct} 
H_{\lbrace 14 \rbrace i}\otimes H_{\lbrace 14 \rbrace j}=
\lbrace H_{\lbrace 8\rbrace i}\oplus H_{\lbrace 3\rbrace i} \oplus H_{\lbrace \bar{3}\rbrace i} \rbrace  \otimes \lbrace H_{\lbrace 8\rbrace j}\oplus H_{\lbrace 3\rbrace j} \oplus H_{\lbrace \bar{3}\rbrace j} \rbrace
\end{multline}
By keeping the possible matrix product from the above equation, the following equality is obtained, 
\begin{equation}\label{14cartanproduct3}
Tr \lbrace H_{\lbrace 14 \rbrace i}\otimes H_{\lbrace 14 \rbrace j}\rbrace =Tr\lbrace H_{8i}H_{8j} \oplus H_{3i}H_{3j}\oplus
H_{\bar{3}i}H_{\bar{3}j}\rbrace ,\nonumber
\end{equation}
\begin{equation}
Tr \lbrace H_{\lbrace 14 \rbrace i} \otimes H_{\lbrace 14 \rbrace j}\rbrace =C^{2}_{\lbrace 8\rbrace}\frac{d_8}{N^2-1}\oplus 2C^{2}_{\lbrace 3\rbrace}\frac{d_3}{N^2-1} 
 =3 \frac{8}{8}+2\frac{4}{3}\frac{3}{8}
 =4.
\end{equation}
\begin{figure*}[!]
\begin{center}
\resizebox{1\textwidth}{!}{
\includegraphics{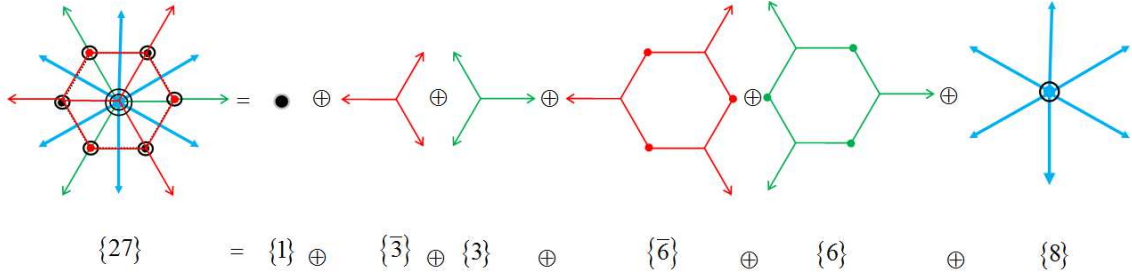}}
\caption{
The $27$ weight diagram of the the $G(2)$ group and its decomposition to the $SU(3)$ group in the weight space.}
\label{weight27}
\end{center}
\end{figure*}  
 The $G(2)$ $27$ weight diagram and its decomposition to its $SU(3)$ subgroup is shown in figure \ref{weight27}. Again it leads to
\begin{equation}
\lbrace 27\rbrace=\lbrace 1\rbrace +\lbrace 3\rbrace+\lbrace \bar{3}\rbrace +\lbrace 6\rbrace +\lbrace \bar{6}\rbrace+\lbrace 8\rbrace ,\nonumber
\end{equation}
\begin{equation}
H_{\lbrace 27\rbrace}= H_{\lbrace 1\rbrace}+H_{\lbrace 3\rbrace}+H_{\lbrace \bar{3}\rbrace}+H_{\lbrace 6\rbrace}+H_{\lbrace \bar{6}\rbrace}+H_{\lbrace 8\rbrace},
\end{equation}
Multiplying two Cartan generators 
\begin{equation}
Tr \lbrace H_{\lbrace 27\rbrace i} \otimes H_{\lbrace 27\rbrace j} \rbrace =
Tr\lbrace H_{8i}H_{8j} \oplus H_{6i}H_{6j} \oplus H_{\bar{6}i}H_{\bar{6}j}
 \oplus H_{3i}H_{3j}\oplus H_{\bar{3}i}H_{\bar{3}j} \oplus H_{1i}H_{1j} \rbrace ,
\end{equation}  
and following the same calculation as the previous it leads to 
\begin{equation}
Tr \lbrace H_{\lbrace 27\rbrace i} \otimes H_{\lbrace 27\rbrace j} \rbrace =
0 +2C^{2}_{3}\frac{d_{3}}{N^{2}-1}+ 2C^{2}_{6}\frac{d_{6}}{N^{2}-1}+C^{2}_{8}\frac{d_{8}}{N^{2}-1}
=0+2\frac{4}{3}\frac{3}{8}+2\frac{10}{3}\frac{6}{8}+3\frac{8}{8}
=9.
\end{equation} 
Now considering the dimension $7$ for the fundemental representation and $14$ for the adjoint representation, the true Casimir ratios for this group are obtained as
\begin{equation} 
1=7C_{7}/48 ,4=14C_{14}/48, 9=27C_{27}/48 \Longrightarrow C_{14}/C_{7}=2 ,C_{27}/C_{7}=\frac{7}{3}=2.33.
\end{equation}  
The results are the true Casimir ratios which can be obtained with the proceture introduced in \cite{olejnk2008, olejnk2008pos} and is applied for Casimir scaling calculation in \cite{Englefield1980}. All of the other higher representations of $G(2)$ can be obtained with the same precedure. Here it is shown that the Casimir scaling behavior within the $G(2)$ decomposition to its subgroup is obtainable if one consider $\alpha (x)=constant $. The Casimir eigen value of $G(2)$ group can be obtained using its decomposition to its $SU(3)$ subgroup. The conclusion here is that the Casimir scaling law of $G(2)$ group can be available for its decomposition to $SU(3)$ group analytically.
Also due to the second assumption of the model the Cartan sub algebra should be used. It should be taken into account this property when decomposition is used for a gauge theory. The Cartan sub algebra for the decomposed sub groups must be taking into account.

In this analytical calculation a constant vortex flux $\alpha_c(x)$ is considered. But this depends on the vortex center relative to the Wilson loop and also the Wilson loop size. To consider such effects the numerical calculation for this model should be done. 
In the next section the confinement potential behavior for the $G(2)$ gauge theory is studied using its decomposition to its $SU(3)$ subgroup without considering the domain structure. 
  
\section{The Confinement Behavior of the $G(2)$ Group Using The Thick Center Vortex Model without the Domain Structures} 
  
To obtain the potential using the thick center vortex model an ansatz for the vortex flux must be used. In this article the one introduced in \cite{Faber1998} is used to obtain the vortex profile $\vec{\alpha}^n_c(x)$:
  
\begin{eqnarray}
\vec{\alpha}^n_c(x)=N^{n}_{i}[1-tanh(ay(x)+\frac{b}{R}],
\end{eqnarray}  
$N^{n}_{i}$ is the normalization number for the vortex type $n$ and $i$ is due to the Cartan generators. $a$ and $b$ are the free constants of the profile, and $y(x)$ is  

\begin{eqnarray}
y(x)=\bigg\lbrace \begin{array}{cc}
{-x}&{ \vert R-x \vert >x}\\{x-R}&{\vert R-x \vert \leq x}
\end{array} ,
\end{eqnarray}

\begin{figure}
\begin{center}
\resizebox{0.5\textwidth}{!}{
\includegraphics{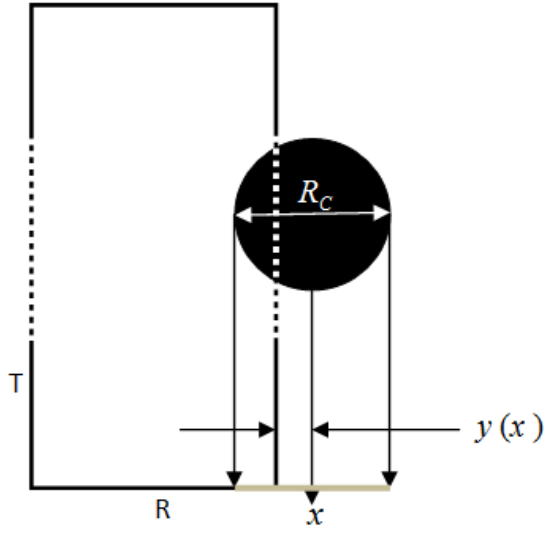}}
\caption{\label{tv} A schematic view of the vortex thickness relative to a planer Wilson loop with  $T\gg R$. $x$ is the position of the center of the vortex. $R_C$ is the vortex thickness and relates to $a, b$. $y(x)$ is the nearest distance to the time like side of the Wilson loop. }

\end{center}
\end{figure} 
  
A schematic view of the vortex thickness relative to a planer Wilson loop with  $T\gg R$ is shown in the figure \ref{tv}. $y(x)$ is the nearest distance of $x$ from time like side of the Wilson loop. $R_C$ is the vortex thickness and relates to $a, b$. The normalization constants $N^{n}_{i}$ are obtained from the maximum flux condition where the loop contains the vortex completely 
\begin{eqnarray}
exp(i\vec{\alpha}^n.\vec{H})=Z_{n}I,
\end{eqnarray}  
With 
\begin{eqnarray}
z_{n}=e^{\frac{2\pi in}{N}}\in Z_{N},
\end{eqnarray} 
And $I$ is the $r \times r$ unit matrix. This rule is valid for the $SU(N)$ gauge theory, but what about the $G(2)$ gauge theory with trivial center element? surely no nontrivial center element means no center vortex. According to the thick center vortex no center vortex means no confinement. So the potential behavior must shows a screening effect. Also lattice gauge simulations shows a screening behavior for the asymptotic region\cite{olejnk2008, olejnk2008pos}. But according to the numerical calculation there is a linear intermediate distance behavior present for the interquark potential. In the previous section it is shown that the second Casimir eigen value and the Casimir scaling behavior  can be obtained using the $G(2)$ decomposition to the $SU(3)$ representations. This means that its $SU(3)$ subgroup content can explain the Casimir scaling region behavior observed in LGT results. This leads to the conjecture that the properties of the Casimir scaling region also can be described with the $SU(3)$ subgroups decomposition. So instead of normalization to trivial center element of the $G(2)$ it is normalized to the $SU(3)$ center vortices in the $G(2)$. So this leads to 
 \begin{eqnarray}
exp(i\vec{\alpha} ^n .\vec{H})=
\begin{bmatrix}
Z_{3} I_{3\times 3}&        0                    &0\\
0                    &          1                  &0\\
0                    &            0                 &Z_{3}^{*}I_{3\times 3},
\end{bmatrix}
\end{eqnarray} 
where $Z_{3}$, $Z_{3}^{*}$ are the nontrivial center element of $SU(3)$ and the $SU(3)^*$ parts. The upper $I_{3\times 3}$ matrix and the lower one leads to similar normalization condition. This means that the maximum flux of the two $SU(3)$ center vortices are the same but the direction is opposite. The presence of two center vortex for applying the thick center vortex leads to interesting situation which will be discussed in the next section.  If a Wilson loop is large enough to encompass both the vortices, then the total difference due to these two vortices is zero. This is because 
\begin{eqnarray}
\langle W(C)\rangle = Tr(
\begin{bmatrix}
Z_{3} I_{3\times 3}&        0                    &0\\
0                    &          1                  &0\\
0                    &            0                 & Z_{3}^{*}I_{3\times 3}
\end{bmatrix}
UUU\dots U) \nonumber
\end{eqnarray}
\begin{eqnarray}
\langle W(C)\rangle =e^{\frac{2\pi i n}{3}} Tr(
\begin{bmatrix}
 I_{3\times 3}&        0                    &0\\
0                    &          e^{\frac{-2\pi i n}{3}}                  &0\\
0                    &            0                 & e^{\frac{-2\pi i n}{3}}Z_{3}^{*}I_{3\times 3}
\end{bmatrix}
UUU\dots U )\nonumber
\end{eqnarray}
\begin{eqnarray}
e^{\frac{2\pi i n}{3}}e^{\frac{-2\pi i n}{3}} Tr(
\begin{bmatrix}
Z_{3} I_{3\times 3}&        0                    &0\\
0                    &          1                 &0\\
0                    &            0                 & Z_{3}^{*}I_{3\times 3}
\end{bmatrix}
UUU\dots U )
=e^{\frac{2\pi i n}{3}}e^{\frac{-2\pi i n}{3}}\langle W(C)\rangle. 
\end{eqnarray} 
Here the effects of the two $SU(3)$ and $SU(3)^*$ center vortices on the Wilson loop are considered. First the $SU(3)$ content of presence of the center vortex is extracted. Second the $SU(3)^*$ content of presence of the vortex is extracted. Then these two parts can eleminate each other and lead to the former Wilson loop. So presence of the two $SU(3)$ vortices with opposite flux for the $G(2)$ leads to the condition that no holonomy is changed for the large Wilson loop. But if the Wilson loop is not large enough it does not encompass the two vortices totally and the effects of two vortices are not trivial on the Wilson loop. To consider the effects of presence of the two $SU(3)$ vortices with opposite flux, the vortex flux should be normalized to $e^{2\pi i}$. The maximum value of $SU(3)$ and $SU(3)^*$ vortices embedding in the $G(2)$ group is obtained with normalization to $e^{2\pi i}$. For the normalization it is obtained: 

\begin{equation}
\alpha^{max}_{3}=0,
\alpha^{max}_{8}=2\pi \sqrt{24}.
\end{equation} 

 \begin{figure}[t]
\begin{center}
\resizebox{1\textwidth}{!}{\includegraphics{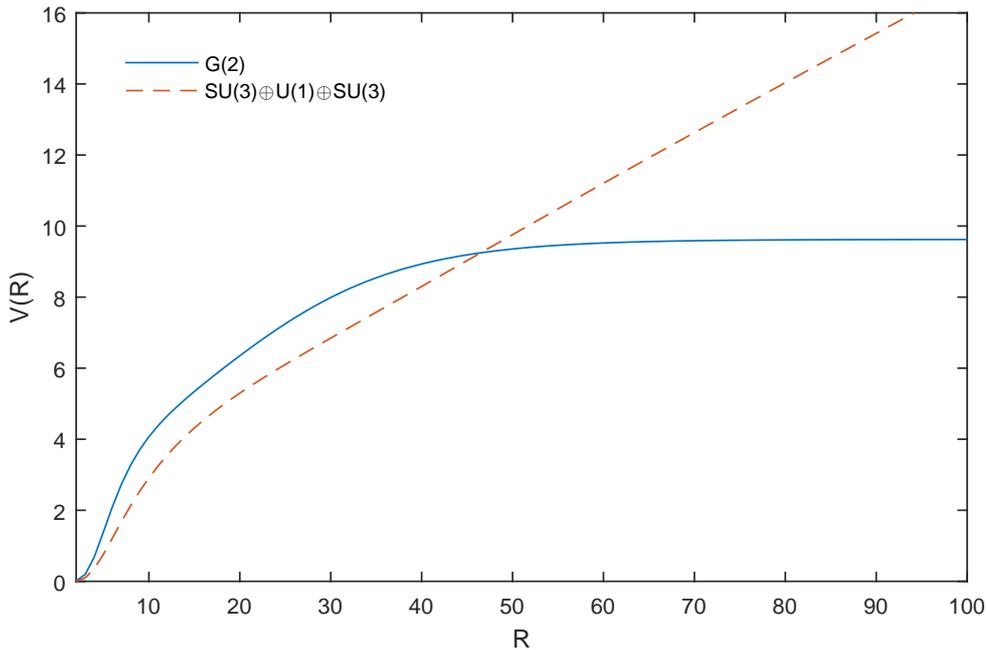}}

\caption{ 
The potential $V(R)$ for the static quark-antiquark sources with the gauge symmetry of $G(2)$ in the fundamental representation using the thick center vortex model without domain structure. The model is applied to the internal $SU(3)$ vortices. Presence of the two $SU(3)$ vortices with opposite fluxes only annihilate each other when the Wilson loops are large and the screening effect happens. Also the potential of a $SU(3) \oplus U(1) \oplus SU(3)$ symmetry is plotted. The second $G(2)$ slope is mainly the slope of $3-ality$ region of the $SU(3) \oplus U(1) \oplus SU(3)$ symmetry.}
\label{fundrep}
\end{center}
\end{figure}

\begin{figure}[t]
\begin{center}
\resizebox{0.8\textwidth}{!}{
\includegraphics{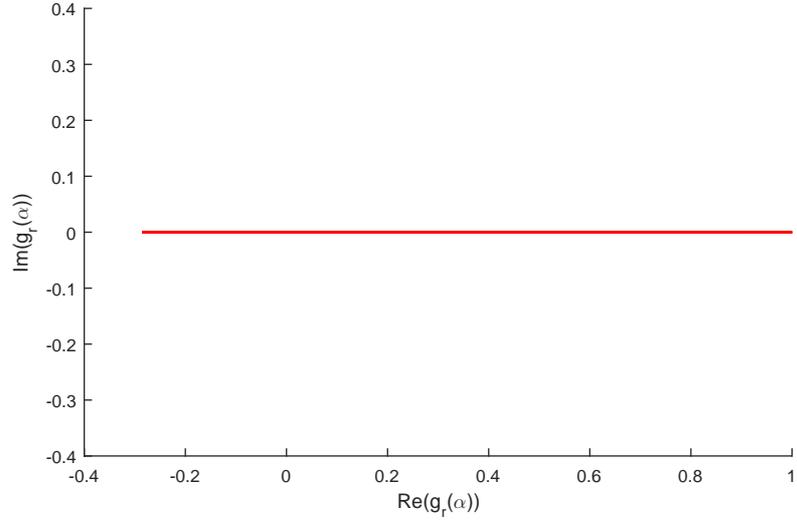}}

\caption{The imaginary part compared to the real value of $g_{r}(\alpha(x))$ for the $G(2)$ gauge theory. The elements of the $G(2)$ are real and no imaginary part is present. The real value of $g_{r}(\alpha(x))$ have values from $1$ to $-0.28$. The minimum can be explain with the $SU(3)$ center elements. So the $SU(3)$ vortices explain the value $-0.28$. }
\label{realimageg2}
\end{center}
\end{figure}

\begin{figure}[!]
\resizebox{0.8\textwidth}{!}{
\includegraphics{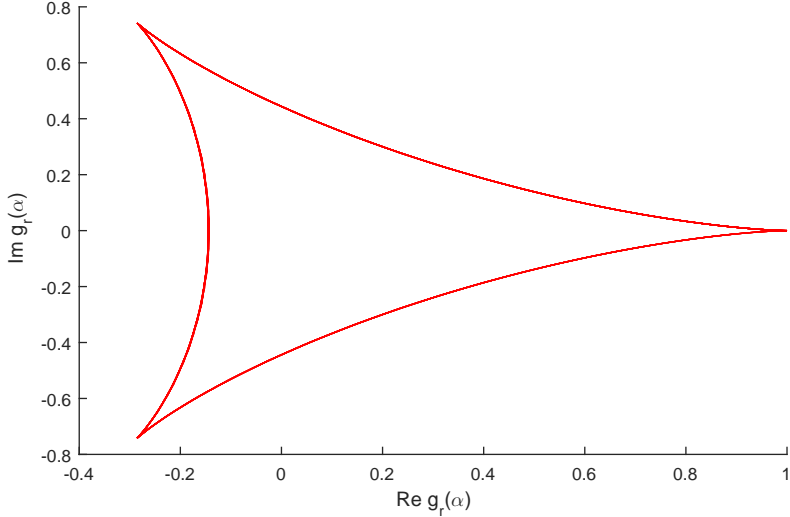}}

\caption{
Instead of the $G(2)$ group the imaginary compared to the real value for a symmetry with $SU(3)\oplus U(1)\oplus SU(3)$ structure is considered here. This leads to the presence of the imaginary parts. The imaginary values of this figure are zero for $1$ and $-0.14$ which can explain the maximums of $Re g_{r}(\alpha(x))$ at $1$ and $-0.14$.}
\label{realimageg2a}
\end{figure}

\begin{figure}[!]
\begin{center}
\resizebox{0.8\textwidth}{!}{\includegraphics{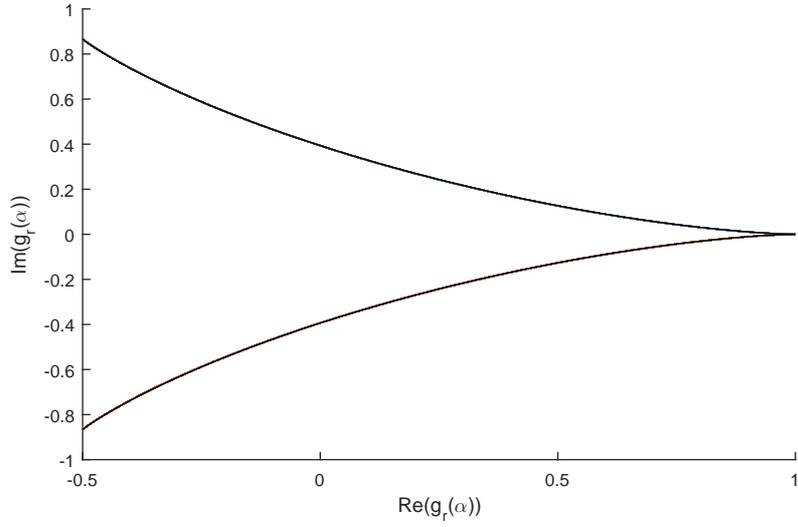}}

\caption{
The imaginary value compared to the real value of $Re g_{r}(\alpha(x))$ for the $SU(3)$ group. The real values change from $-0.5$ to $1$. $-0.5$ is the nontrivial center element value for this group.} 
\label{realimagesu3}
\end{center}
\end{figure}

\begin{figure*}[t]
\resizebox{1\textwidth}{!}{
\includegraphics{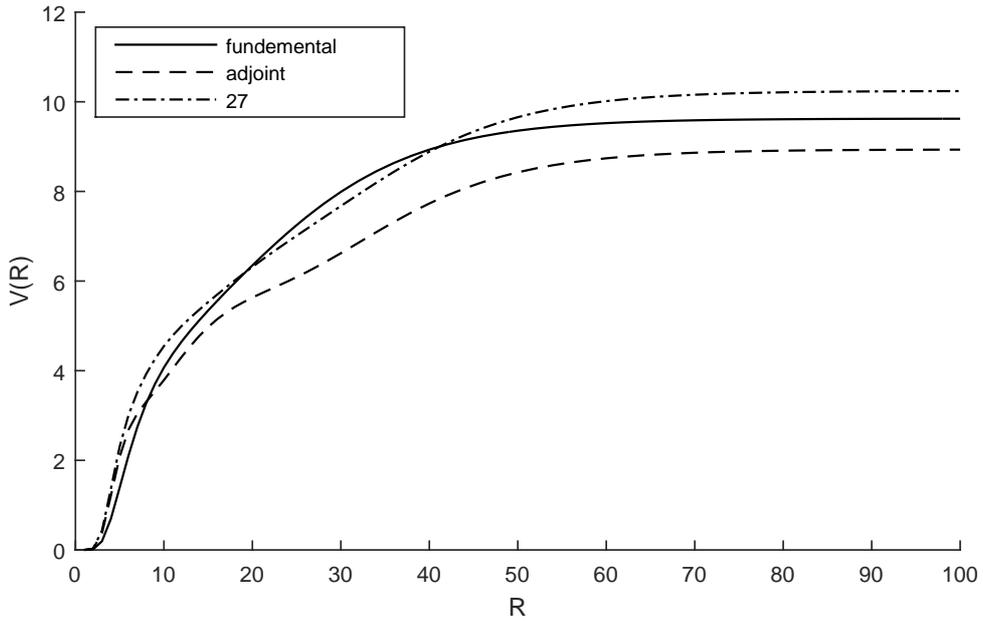}}

\caption{ The potential of static sources of quark anti quark with a $G(2)$ gauge theory in the fundamental and adjoint and the $27$ representations. The thick center vortex model is applied to the internal $SU(3)$ vortices. The behavior of confinement in the Casimir region and the asymptotic region can be explained using these internal vortices without considering the domain structure.} 
\label{potentialall}
\end{figure*}

\begin{figure}[!]
\resizebox{1\textwidth}{!}{\includegraphics{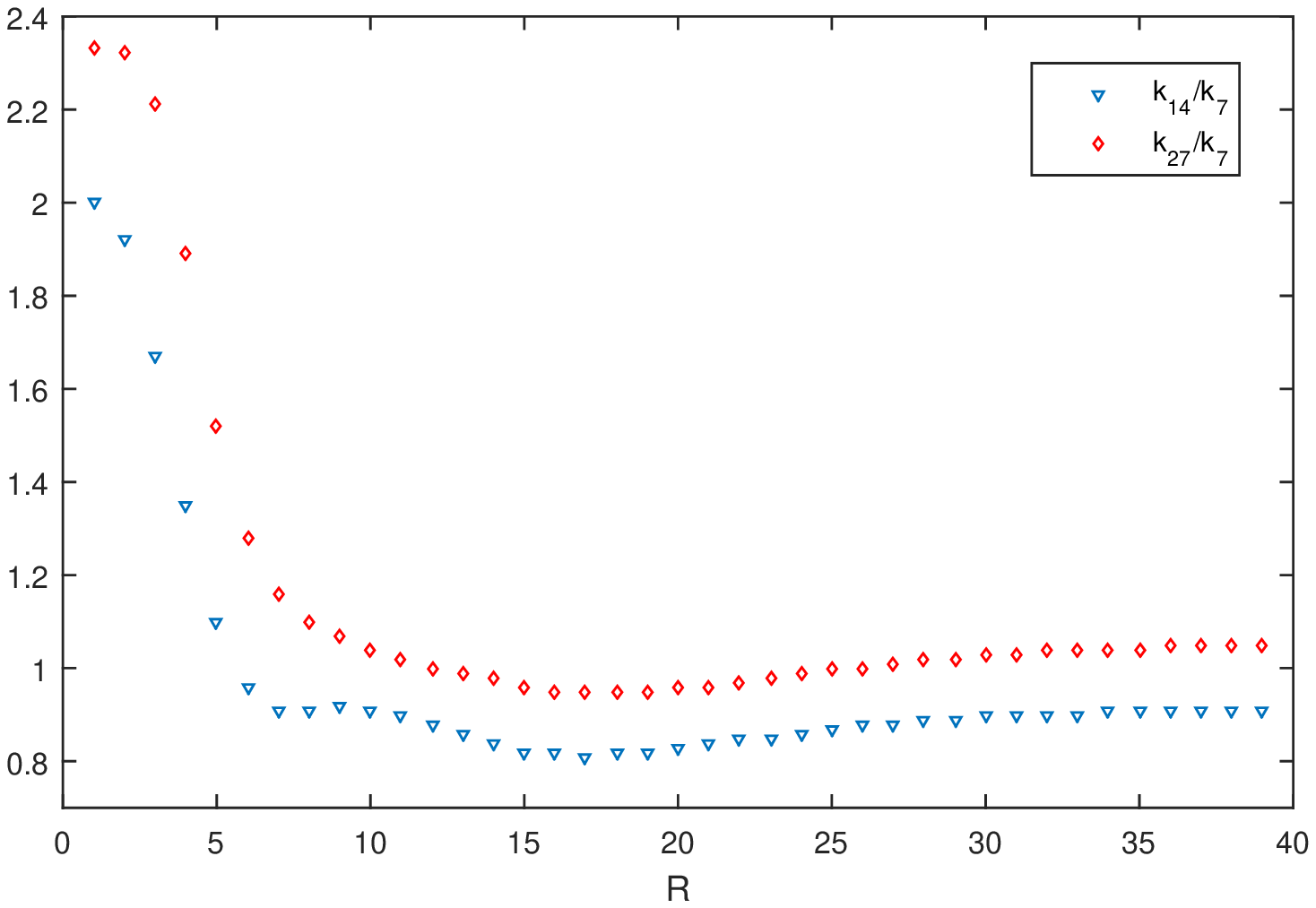}}
\caption{The potential ratios for $k_{14}/k_{7}$ and $k_{27}/k_{7}$. Upper than $43$ percent Casimir scaling is observed.} 
\label{ratio}
\end{figure}

Using these normalization condition the potentail for the fundemental representation of the $G(2)$ gauge theory can be obtaind. The potentail is obtained by the following formula
\begin{equation}
V(R)=\sum_{x} ln\lbrace 1- \sum_{n=1}^{N-1} f_{n} (1- Re g_{r}[\vec{\alpha}_{C}^{n}(x)])\rbrace .
\end{equation}
The free parameters of the model are considered as $a=0.05, b=4, f=0.1$ for the present numerical calculation. Figure \ref{fundrep} shows the fundemental potentail for the fundemental representation of the $G(2)$ group using the thick center vortex without modification to the domain structure. Here a $f_{0}$ value for the probability of considering the trivial domain is not considered. Instead, the thick center vortex model is applied for the $SU(3)$ subgroups of the $G(2)$. Also an additional assumption of the vortex flux direction is considered. As it is clear in the figure, an asymptotic screening behavior similar to what is observed using lattice gauge theory \cite{olejnk2008}, is obtained. Also an intermediate linear potential is observerd. The potential of a $SU(3)\oplus U(1)\oplus SU(3)$ symmetry also is plotted. The second $G(2)$ slope is mainly the slope of $3-ality$ region of the $SU(3)\oplus U(1)\oplus SU(3)$ symmetry. However due to the final slope of the potentails they differ from each other slightly. The real value compared to the imaginary part of the $g_{r}(\alpha)$ is plotted for $G(2)$ in the figure \ref{realimageg2}. In the figure \ref{realimageg2a} this ratio is plotted considering the two $SU(3)$ subgroups of the $G(2)$ with positivie flux for both of the subgroup instead of considering the two subgroups with true fluxes which annihilate the imaginary part of each other as in the figure \ref{realimageg2}. This figure help us to understand the behavior of $Re g_{r}(\alpha)$ and its maximum and minimum. The same figure is plotted for the $SU(3)$ group in the figure \ref{realimagesu3}. In this figure the maximum is $1$ and the minimum is $-0.5$ which is due to the maximum value of the flux for this group. This value is obtained by setting $n=0$ in $e^{\frac{2\pi i n}{3}}$. The minimum is also obtained by $n=1$. In the $G(2)$ gauge theory also the properties of $g_{r}(\alpha)$ can be explained by its $SU(3)$ subgroups. The absolute minimum is obtained \cite{lookzadeh2012, lookzadeh2011} from the $SU(3)$ minimum as

\begin{equation} \label{minmax}
Re g_{r}(\alpha)_{min}=\frac{1}{7}Tr(e^{i\alpha H})_{min}=  
\frac{1}{7} Tr
\begin{bmatrix}
g^\prime_{R}(\alpha)_{min}\times d_{r} &  0  & 0 \\
0                                                    &  1  &  0\\
0                                                    &  0  & g^\prime_{R}(\alpha)_{min}\times d_{r}
\end{bmatrix}\nonumber
\end{equation}
\begin{equation}
Re g_{r}(\alpha)_{min}==\frac{1}{7}(-0.5\times 3+1-0.5\times 3)=- 0.28. 
\end{equation} 

 Which $ g^{'}_{R}(\alpha)_{min}$ is the minimum value of the flux in the $SU(3)$ subgroup. Despite the maximum and the minimum of the $Re g_{r}(\alpha)$ can be obtained using the above equation \ref{minmax} the local minimum and maximum of $Re g_{r}(\alpha)$ can be exist. In the figure \ref{grg2} the $Re g_{r}(\alpha)$ is shown for the fundemental representation and for the different Wilson loop distances. These local maximums are occured at  $-0.14$. To obtain this extremum instead of the $G(2)$, a group is considered which has the properties of the $G(2)$ decomposition except $\bar{3} \rightarrow 3$, which leads to the two $SU(3)$ vortices with positive flux. Then the behavior of the real compared to the imaginary part of $g_{r}(\alpha)$ for such symmetry is plotted in the figure \ref{realimageg2a}. The maximum in this figure occur where the imaginary of $g_{r}(\alpha)$ become zero. So such solution for the imaginary part of the $g_{r}(\alpha)$ is founded as

\begin{multline}
Im(g_{r}(\alpha))=\frac{1}{7} Im(e^{i\vec{\alpha} .\vec{H}})=
(1/7)(sin(H_{(1,1)}\alpha)+sin(H_{(2,2)}\alpha)+sin(H_{(3,3)}\alpha)+sin(H_{(4,4)}\alpha)\\
+sin(H_{(5,5)}\alpha)+sin(H_{(6,6)}\alpha)+sin(H_{(7,7)}\alpha))=0
2sin(\alpha)+sin(-2\alpha)=0.
\end{multline}  
The $H_{(a,a)}$s are the $(a,a)$ component of the $H^8$ diagonal Cartan generator. The solutions are $0, \pi, 2\pi, \dots $. Putting the $\alpha=\pi$ in the real part of $g_{r}(\alpha)$
\begin{equation}
Re(g_{r}(\alpha))=(1/7) Re(e^{i\vec{\alpha} .\vec{H}})=
(1/7) (2(cos(\alpha)+cos(-\alpha))+cos(2\alpha)+cos(-2\alpha)+1)\nonumber
\end{equation}
\begin{equation}
(1/7)(2(-1-1)+1+1+1)=-0.14.
\end{equation}
So the maximum value is obtained equal to the value obtained in the figure \ref{grg2}. Another explanation is that the extremum occur when $\frac{d Reg_{r}(\alpha)}{d\alpha}=0$. This leads to
\begin{equation}
dRe(g_{r}(\alpha))/d\alpha=(1/7) dRe(e^{i\vec{\alpha} .\vec{H}})/d\alpha \nonumber
\end{equation}
\begin{equation}
-2(sin(\alpha)-sin(-\alpha))-2sin(2\alpha)+sin(-2\alpha)=0.
\end{equation}
So again the same solutions $0,\pi, 2\pi $ which leads to the $-0.14$ is obtained.
   
To have a judgement for the Casimir scaling region the potential behavior between the two static quark anti quark in the fundamental and the adjoint and the $27$ representations are obtained. The normalization condition for the higher representations are also obtained using the $SU(3)$ decompositions of the higher representations of the $G(2)$ group. Figure \ref{potentialall} shows the potential for the higher representations. The Casimir scaling are $c_{14}/c_{7}=2, c_{27}/c_{7}=2.33$ for the adjoint and $27$ representation. 
 The potential ratio for $k_{14}/k_{7}$ and $k_{27}/k_{7}$ are represented in the figure \ref{ratio}. Upper than $43$ percent Casimir scaling is observed. Other studies for the Casimir scaling also show such behavior within the thick center vortex model \cite{Deldar2001, Deldar2005, Rafibakhsh2007, Rafibakhsh2010, Lookzadeh2012}. Also the screening is due to the zero total flux of two "internal vortices" with the opposite direction. 
  
\section{The internal $SU(3)$ Thick Center Vortex within the $G(2)$ Gauge Theory }
As it can be seen in the previous sections, using the thick center vortex for the $SU(3)$ decomposition of the $G(2)$, one can obtain the true $N$-ality and the true Casimir scaling behavior without the domain structure consideration. Whenever the Wilson loop is large enough to encompass the whole vortices, which means the two internal $SU(3)$ and $SU(3)^*$ vortices with different direction of fluxes, these two vortices lead to the zero center flux. Zero center vortex flux means no whole vortex present and screening behavior at asymptotic region must be observed. So considering the $N$-alities for the $SU(3)$ subgroups leads to zero-ality and the screening effect must be observed. When the Wilson loop does not encompass the whole two vortices, then the net two vortex fluxes are not zero and these $SU(3)$ vortices flux lead to the confinement behavior at intermediate distance in accord with an acceptable Casimir scaling behavior. 

\begin{figure*}[t]
\begin{center}
\resizebox{1\textwidth}{!}{
\includegraphics{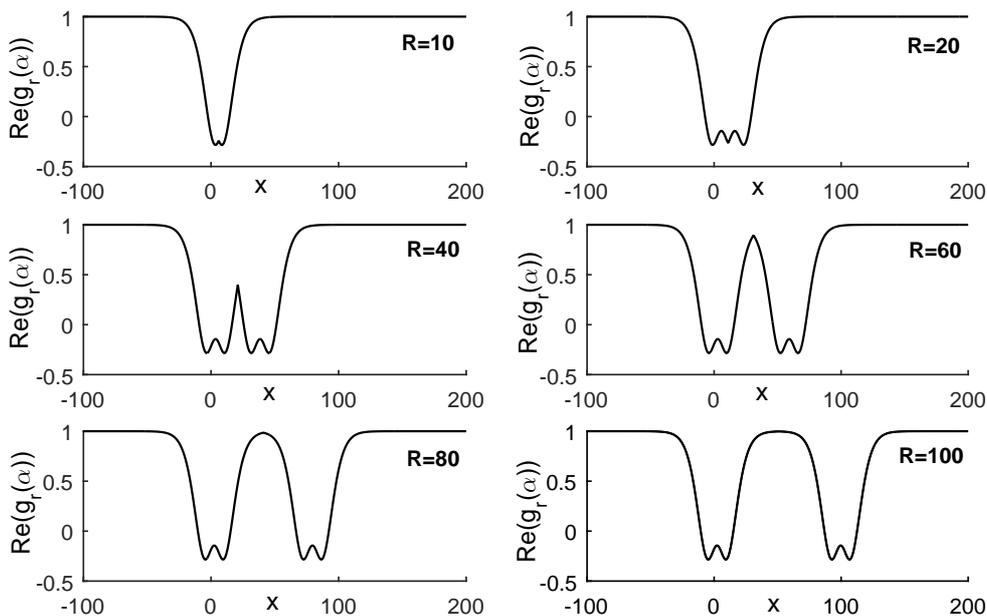}}
\caption{
The plots of $Re g_{r}(\alpha(x))$ for the fundamental representation of the $G(2)$ group at different Wilson loop values. The $-0.28$ minimum within these figures can be explained with the $SU(3)$ nontrivial center elements. The maximum values at $-0.14$ are also due to the zeros of $dg(\alpha)/d\alpha$. These figures at all distances have two symmetric parts. At the distances $R=10,20,40,60$ the profiles are changing and this leads to the linear part of the potential. However for $R=80,100$ the profiles do not changed and only the distances between the two symmetric parts are increased. This leads to a zero potential slope and the screening effect behavior. At such distances the Wilson loop is large enough to encompass the two $SU(3)$ vortices with the opposite fluxes.}
\label{grg2}
\end{center}
\end{figure*}

To understand such behavior better, the $Re g_{r}(\alpha(x))$ is plotted for the different lengths $R$ of the Wilson loops in the figure \ref{grg2}. Considering the potential figure \ref{potentialall}, it is observed that at the distances $R=10$ to $60$ the potentials have the linear behavior and from the $60$ to $100$ the potentials have zero slope. This can be explained using the $Re g_{r}(\alpha(x))$ behavior.
The $Re g_{r}(\alpha(x))$ is the effect of the two $SU(3)$ and $SU(3)^*$ vortices as their positions are moved in the presence of the Wilson lines. The two vortices of $SU(3)$ and $SU(3)^*$ are positioned at the same place and can not move separately. Also there are two symmetric parts in these figures. These two symmetric parts arise due to the symmetric effect of the vortices relative to the left and right time like sides of the Wilson loop. In the figure \ref{tv} both of the cross sections to the Wilson loop from right and left lead to a similar effect on the Wilson loop. These two symmetric parts are due to this behavior. When the Wilson loop is smaller than the vortex thickness, the effect of vortices flux on the $Re g_{r}(\alpha(x))$ are not complete. If only one of the symmetric parts is considered, it is seen that this part of the figure is not symmetric due to its minimum or local maximum value at $-0.14$. By increasing the Wilson loop $R$ size the Wilson loop becomes large enough to have the whole vortices flux. This leads to two symmetric parts in the $Re g_{r}(\alpha(x))$ within each parts are symmetric relative to its local maximum value at $-0.14$. This leads to the complete two $SU(3)$ and $SU(3)^*$ vortices parts present in the Wilson loop and then no vortex situation. No variation in the shape of the two symmetric parts leads to the no variation in the potential behavior and the screening behavior happens.

When the Wilson loop is large enough to have the both symmetric parts of the figure relative to their local maximum, it means that the two $SU(3)$ and $SU(3)^*$ vortices holonomy completely link the Wilson loop instead of partial linking occur due to the vortex thickness. The two vortices with complete and opposite flux annihilate each other and lead to the no vortex situation at large distances. But at smaller distances the $SU(3)$ vortices flux parts don't annihilate the effect of each other. So a non-zero slope region is observed and the Casimir scaling can be obtained roughly as it is obtained in the previous section.   

According to the thick center vortex the vortices flux fluctuate independently relative to each other. But here it is considered two $SU(3)$ vortices effects on the confinement. Are these internal vortices flux independent or they can interact with each other? The simple situation is the one in which there is no interaction between these two vortices. Another interesting situation is to study the interaction between these two vortices. The thick center vortex is introduced in $D=2$ slice of $D=4$ space time. In the reference \cite{Greensite2007} it is shown that the Casimir behavior can be obtained in $D=2$ slice of $D=4$ gauge theory. So the thick center vortex model is introduced in $D=2$ dimension to explain the Casimir scaling behavior. The same behavior can take place for the plane parallel to this plane which make the thick center vortex a line like object in $D=3$ dimension. Considering the time, it leads to a surface like object. 

If the vortex free energy density is known, it is possible to put the vortices at different distances and through the energy of the two vortices at different distances the interaction between vortices can be studied \cite{lookzadeh2012,lookzadeh2014}, However in this model the portion of the linking of a vortex with a Wilson loop is considered and the whole center vortex profile is not considered at any Wilson distances. Despite this, one can study the effects of considering two center vortices portion at different distances on the center vortex by considering the vortices flux which are localized at two points with the known distance between them and study the effect of such vortices on the Wilson loops. These vortices must have different properties relative to the vortices explained here, because the $SU(3)$ and $SU(3)^*$ vortices explained here can not be separated relative to each other. However here only two $SU(3)$ vortices with the same types of fluxes at the same place with no interaction between them is considered. The two $SU(3)$ and $SU(3)^*$ vortices holonomy linking with Wilson loops are complete for the large loop and they are partial when the Wilson loops are small. Then only for the large Wilson loops the two vortices annihilate each other effect on the Wilson loop completely and a screening behavior at large distance is observed.  
\section{Conclusion }  
Using the thick center vortex model it is tried to understand the behavior of the confinement of the $G(2)$ gauge theory without modification to the domain model. It is shown that what would be happened if the Cartan sub algebra is used within the thick center vortex. The Casimir scaling can be obtained analytically. Also according to the second assumption of the thick center vortex model, the Cartan sub algebra must be used to have the most abelian part present in the calculation. The Casimir ratios of this group are obtained using its decomposition to the $SU(3)$ groups analytically. This leads to the conjecture that the properties of confinement can be obtained by these $SU(3)$ subgroups. Also it would be interesting if the Casimir ratios can be obtained analytically using the $SU(2)$ subgroups of the $G(2)$ gauge group. Then the thick center vortex model for the $SU(3)$ subgroup of $G(2)$ is used due to the presence of a $SU(3)$ and a $SU(3)^*$ part. The two $SU(3)$ vortices at the same place with opposite fluxes are considered. When the Wilson loops are large enough to encompass the whole two vortices the two $SU(3)$ vortices annihilate each other and lead to the no vortex situation for the large Wilson loops. This leads to true $0- ality$ and screening effect. However when the Wilson loop is small the effects of the two $SU(3)$ and $SU(3)^*$ vortices on the Wilson loop are present and lead to the linear behavior with an acceptable Casimir scaling behavior within the thick center vortex model. The second slope of the potential at the intermediate distances is slightly equal to the slope of the $3-ality$ part of the potential at asymptotic distances in a $SU(3)+U(1)+SU(3)$ symmetry. The minimum and maximum of the $Re g_{r}(\alpha(x))$ is also obtained using its decomposition to the $SU(3)$ subgroups. The minimums are due to the maximum flux of the $SU(3)$ vortices and the maximums are due to the zeros of $\frac{d Re g_{r}(\alpha)}{d\alpha}$. The two $SU(3)$ vortices here are considered, have similar fluxes and are positioned at the same place. No interaction is considered for these types of vortices embedding in the $G(2)$ gauge theory. Using the idea of domains for the vacuum and different types of fluxes, one can modify the idea of thick center vortex and also a broader Casimir region can be obtained. Here the old ansatz and idea of thick center vortex is used to obtain the properties of the confinement in the $G(2)$ gauge theory with the trivial center element. According to this choice, the properties of the confinement within such gauge theories governed by their ''internal vortices'' due to the possiblity of decomposition of these gauge theories to the groups with the non trivial center.  

The $G(2)$ gauge theory has interesting symmetries which make it a good mathematical laboratory for the better understanding of the physics of the $SU(N)$ gauge theories. Here due to the trivial center element of the $G(2)$ theory, properties of the confinement is studied using the thick center vortex model and the $G(2)$ decomposition to the $SU(3)$ group. The lattice simulation is not dependent on the algebra which is used for generating the gauge group elements. Despite this it would be interesting if other properties within the $G(2)$ group can be investigated by such kind of the decomposition algebra and take a control on the properties of the $SU(3)$ in the simulation \cite{Gattringer2009,Lucini2006}. Due to this explicit $SU(3)$ subgroup, lattice simulation of a $G(2)$ theory is straightforward but is expensive \cite{Wellegehausen2012}. $G(2)$ thermodynamics properties and the gluon propagator and the running coupling properties are more similar to the $SU(3)$ group\cite{MaasWellegehausen}. The confining/deconfining mechanism is common to all non-Abelian theories, irrespective of the underlying gauge group. The order of the deconfining transition, however, does depend on the gauge group: for the $SU(2)$ case, the mechanism predicts the existence of a second-order transition, whereas for the $SU(N \geq 3)$ and for the $G(2)$ the transition is a discontinuous one \cite{Caselle2015}. It may be interesting to study $G(2)$ lattice simulation from such $SU(3)$ decomposition. Also Abelian monopoles properties within the $G(2)$ gauge theory using the $SU(3)$ decomposition can be examined \cite{Lucini2006,Di Giacomo2007}. The possibility of existence of QCD matter in a neutron star core with the $G(2)$ symmetry also seems interesting. Due to the decomposition here explained it is possible to compare such matter with the usual QCD matter \cite{Maas2016, Maas20161}. 

{\bf ACKNOWLEDGMENT}
 
I would like to thank Pro. Gernot Munester, Pro. Marco Panero, Dr. Shahnoosh Rafibakhsh for their helpful comments.

\end{document}